\documentclass[aps,pra,preprint,groupedaddress,showpacs]{revtex4}
\usepackage{graphicx}
\usepackage[ansinew]{inputenc}
\usepackage{array}
\usepackage{color}
\usepackage{amsmath}
\usepackage{amsxtra}
\usepackage{amssymb}
\usepackage{latexsym}
\usepackage{dsfont}
\begin{document}
\title{Entanglement of a Laguerre-Gaussian cavity mode with
a rotating mirror}
\author{M. Bhattacharya, P.-L. Giscard, and P. Meystre}
\affiliation{B2 Institute, Department of Physics and College of
Optical Sciences, The University of Arizona, Tucson, Arizona
85721}

\date{\today}

\begin{abstract}
It has previously been shown theoretically that the exchange of
linear momentum between the light field in an optical cavity and a
vibrating end mirror can entangle the electromagnetic field with
the vibrational motion of that mirror. In this paper we consider
the rotational analog of this situation and show that radiation
torque can similarly entangle a Laguerre-Gaussian cavity mode with
a rotating end mirror. We examine the mirror-field entanglement as
a function of ambient temperature, radiation detuning and orbital
angular momentum carried by the cavity mode.
\end{abstract}


\maketitle

\section{Introduction}
\label{sec:intro} The optomechanical driving of optical cavities
with laser radiation is an experimentally promising method for
eliciting quantum mechanical behavior from classical objects
\cite{gigan2006,kleckner2006,arcizet2006,schliesser2006,corbitt2007}.
In the case of a two-mirror cavity with a perfectly reflecting
end-mirror mounted on a cantilever and allowed to vibrate along
the cavity axis, laser radiation entering through the slightly
transmissive fixed input mirror can both trap and cool the mirror
motion, depending on whether it is detuned slightly above or below
a cavity resonance \cite{corbitt2007}. Radiation pressure can thus
decrease the number of phonons in the vibrating mirror,
\begin{equation}
\label{eq:nosc}
n_{m} = \frac{k_{B}T_{\rm eff}}{\hbar \omega_{\rm eff}},
\end{equation}
since cooling lowers the effective temperature of the mirror
$T_{\rm eff}$ while trapping increases its effective vibration
frequency $\omega_{\rm eff}$. A key goal is to achieve $n_{m} <1$,
i.e. to place the mirror near its quantum mechanical ground state.
Evidence of reaching that regime could come from measuring
$n_{m}$, and a strategy for doing so has recently been proposed
\cite{thompson2007}.

Radiation pressure can play an additional important role as a
mechanism for generating entanglement. In addition to its basic
interest, quantum entanglement - the existence of correlations
disallowed by classical physics between two objects - has been
demonstrated to be a valuable resource for quantum information
processing tasks such as teleportation and superdense coding (see
Ref.~\cite{Nielsenbook} and references therein).

It has been shown theoretically that radiation pressure can
entangle a light field to a vibrating mirror both in the absence
\cite{pirandola2003} and in the presence \cite{vitali2007} of a
cavity. It can also entangle two vibrating mirrors, both when they
form a cavity \cite{mancini2002,pinard2005} and when they do not
\cite{braunstein2003,pirandola2006}. A single vibrating mirror can
also entangle two \cite{bose1997,giovanetti2001} or more
\cite{mancini2001} light fields. The teleportation of quantum
states from a light field to a vibrating mirror has been proposed
in Ref.~\cite{mancini2003}.

In these proposals the physical mechanism behind the generation of
entanglement is the exchange of \textit{linear} momentum between
the photons in the cavity and the mirror. In recent work, we have
proposed a rotational analog of mirror cooling and trapping where
the light-mirror interaction relies on the exchange of
\textit{angular} momentum between a Laguerre-Gaussian cavity mode
and a spiral phase element used as a cavity end-mirror
\cite{mishpm2}. This article further investigates that system and
shows how the radiation torque can generate entanglement between
the optically charged light field and a rotating mirror. We
investigate the entanglement as a function of ambient temperature,
detuning of the laser radiation, and orbital angular momentum
carried by the cavity mode.

The rest of the article is organized as follows.
Section~\ref{sec:cavitymodel} briefly reviews the cavity
configuration proposed in Ref.~\cite{mishpm2}.
Section~\ref{sec:qle} discusses the steady state of the system,
discusses the linear fluctuations of the field and mirror motion
about that steady state, and examines the dynamical stability of
the rotating mirror. Section~ \ref{sec:ent} demonstrates the onset
of entanglement between the intra-cavity Laguerre-Gaussian mode
and the rotating end-mirror and discusses its dependence on
ambient temperature, detuning, and angular momentum of the light
field. We also briefly discuss the experimental measurement of the
entanglement. Section~\ref{sec:conc} provides a summary and a
conclusion. Appendix~\ref{sec:appa} derives and solves the
Lyapunov equation for the correlation matrix determining the
entanglement between the mirror and the cavity mode.

\section{Cavity model}
\label{sec:cavitymodel}

\begin{figure}
\includegraphics[width=0.5 \textwidth]{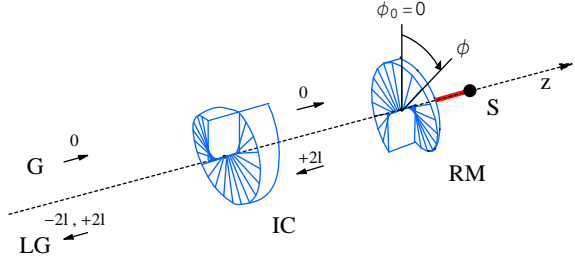}
\caption{\label{fig:RotFieldEntpic1}(Color online). Arrangement
for entangling the rotation of a mirror with a Laguerre-Gaussian
(LG) mode using angular momentum exchange. A Gaussian (G) field of
optical charge 0 enters the cavity through a fixed partially
transparent input coupler (IC) that does not affect its charge. A
perfectly reflecting rear mirror (RM) rotating about the cavity
axis on a support S adds a charge $2l$ to the beam on reflection.
Both the IC and the RM are spiral phase elements. The angular
deflection of the RM from equilibrium $(\phi_{0}=0)$ is indicated
by the angle $\phi$. The charge on the beams at various points is
also indicated.}
\end{figure}

The optomechanical system under consideration is described in
detail in Ref.~\cite{mishpm2} and illustrated in
Fig.~\ref{fig:RotFieldEntpic1}.
Both cavity mirrors are spiral phase elements \cite{thooft1993}.
The input mirror is slightly transmissive but does not change the
orbital angular momentum of any beam passing through it. In
reflection, however, it \textit{removes} a charge $2l$ from the
beam. Likewise the end mirror, which is assumed to be perfectly
reflecting, is designed to \textit{add} a charge $2l$ to a beam on
reflection. With these specifications it can be seen that a
Gaussian beam (with no charge) incident on the cavity can transfer
a torque $\xi_{\phi}=cl\hbar/L$ per photon to the end mirror, where $c$
is the velocity of light and $L$ is the length of the cavity
\cite{mishpm2}.

In dimensionless units, the Hamiltonian that describes this system
is \cite{mishpm2}
\begin{equation}
\label{eq:Ham}
    H = \hbar \omega_{c}a^{\dagger}a+\frac{\hbar\omega_{\phi}}{2}
    \left(L_{z}^{2}+\phi^{2}\right)-\hbar g a^{\dagger}a\phi.
\end{equation}
where $a$ and $a^\dagger$ are the bosonic annihilation and
creation operators for the cavity mode of frequency $\omega_{c}$,
$L_{z}$ and $\phi$ are the angular momentum and angular
displacement, respectively, of the rear mirror about the cavity
axis, with $[L_{z},\phi]=-i$. Finally $\omega_{\phi}$ is the
angular rotation frequency of the rear mirror and
\begin{equation}
\label{eq:optrotcon}
g=\frac{cl}{L}\sqrt{\frac{\hbar}{I\omega_{\phi}}}
\end{equation}
is the optorotational coupling parameter, with $I=M R^{2}/2$ the
moment of inertia of the mirror of mass $M$ and radius $R$ about
the cavity axis.

\section{Quantum dynamics}
\label{sec:qle}

The evolution of the operators describing the dynamics of the
intracavity field and the mirror motion is given by quantum
Langevin equations accounting for noise and damping due to the
vacuum fluctuations entering the cavity field as well as to the
Brownian noise coupling to the moving mirror \cite{mishpm2},
\begin{eqnarray}
\label{eq:QLErot}
\dot{a}&= &-i(\delta-g\phi)a-\frac{\gamma}{2}a+\sqrt{\gamma}a^{\rm in},\nonumber \\
\dot{\phi}&=& \omega_{\phi}L_{z},\nonumber \\
\dot{L_z}&=& -\omega_{\phi}\phi+g a^{\dagger}a
-\frac{D_{\phi}}{I}L_{z}+\epsilon^{\rm in}.
\end{eqnarray}
where $\delta =\omega_{c}-\omega_{L}$ is the detuning of the laser
frequency $\omega_{L}$ from the cavity resonance, $\gamma$ is the
cavity damping rate and $D_{\phi}$ is the intrinsic damping
constant of the rotating mirror. The noise operator $a^{\rm in}$
describes the laser field incident on the cavity, of mean
amplitude $\langle a^{\rm in}(t)\rangle =a_{\rm s}^{\rm in}$. Its
fluctuations are taken to be delta-correlated,
$$
    \langle \delta a^{\rm in}(t) \delta a^{\rm in,\dagger}(t') \rangle=\delta (t-t')
$$
and add vacuum noise to the cavity modes. The Brownian noise
operator $\epsilon^{\rm in}$ describes the mechanical noise that
couples to the mirror from the environment. Its mean value is zero
and its fluctuations are correlated at temperature $T$ as
\cite{gardinerbook}
\begin{eqnarray}
\label{eq:Brownian}
&&\langle \delta \epsilon^{\rm in}(t) \delta \epsilon^{\rm in}(t') \rangle=\nonumber \\
&&\frac{D_{\phi}}{\omega_{\phi}I}\displaystyle\int_{-\infty}^{\infty}
\frac{d\omega}{2\pi}e^{-i\omega(t-t')}\omega \left[1+ \coth
\left(\frac{\hbar \omega}{2k_{\rm B}T}\right)\right],
\end{eqnarray}
where $k_{\rm B}$ is Boltzmann's constant. For high mechanical
quality, $(\omega_{\phi}I/D_{\phi} \gg 1)$, the fluctuations
become delta-correlated and Eq.~(\ref{eq:Brownian}) simplifies to
\begin{equation}
\label{eq:Brownian2}
 \langle \delta \epsilon^{\rm in}(t) \delta
\epsilon^{\rm in}(t') \rangle=
\frac{D_{\phi}}{I}(2\bar{n}+1)\delta(t-t'),
\end{equation}
where
\begin{equation}
\label{eq:nbar} \bar{n}=\left(e^{\hbar \omega_{\rm eff}/k_{B}T}
-1\right)^{-1}
\end{equation}
is the mean number of thermal phonons available at the mirror of
effective rotation frequency $\omega_{\rm eff}$ and temperature
$T$ [see Eq.~(\ref{eq:eff}) for an expresion for $\omega_{\rm eff}$]. 
The thermal phonon spectrum peaks approximately at the temperature
\begin{equation}
\label{eq:Tc} T_{c} \sim \frac{\hbar \omega_{\rm eff}}{4k_{B}},
\end{equation}
We show later on that it is possible to maintain a mirror-field
entanglement close to its maximum value for $T < T_{c}$, but that
that entanglement decreases sharply beyond $T \sim T_{c}$
(Fig.~\ref{fig:RotFieldEntpic3}). This can be understood
intuitively by noting that for $T>T_{c}$ we obtain $\omega>
\omega_{\rm eff}$ and the mirror can be rotationally excited by
thermal phonons, but this ceases to be the case for $T<T_{c}$ , in
which case $\omega < \omega_{\rm eff}$.

\subsection{Steady state}
\label{subsec:ssvalues}

The semiclassical, steady-state solution of Eq.(\ref{eq:QLErot})
is
\begin{eqnarray}
    a_s&=& \frac{\sqrt{\gamma} a_s^{\rm in}[\gamma/2-i(\delta -
    g\phi_s)]}{(\gamma/2)^2+(\delta - g \phi_s)^2} \nonumber \\
    \phi_s&=&\frac{g|a_s|^2}{\omega_\phi},\nonumber \\
    L_{z,s}&=&0.
\end{eqnarray}
The steady state field amplitude $a_s$ is complex in general, but
in simple situations where the incident field is kept constant
throughout, it is possible to chose its phase such that $a_s$ is
real,
\begin{equation}
\label{eq:sstate} a_{\rm s}=\frac{\sqrt{\gamma}|a_{\rm s}^{\rm
in}|} {\left[(\frac{\gamma}{2})^{2}+(\delta-g\phi_{\rm s})^{2}
\right]^{1/2}}.
\end{equation}
We restrict our discussion to that situation here. The
steady-state is bistable for high enough incident field
\cite{pm1985,mancini1994}, but we assume in the following that an
electronic feedback loop maintains the length of the cavity and
hence removes bistability. This is a standard practice in current
experiments on optomechanical cooling
\cite{gigan2006,kleckner2006,arcizet2006,schliesser2006}.

\subsection{Fluctuations}
\label{subsec:flucs}

To account for the quantum fluctuations about the semiclassical
steady-state we expand the Heisenberg operator $a$ about the $a_s$
state in the usual way as $a=a_{s}+\delta a$, and similarly for
the other operators. Linearizing the quantum Langevin equations of
motion in the fluctuations gives then
\begin{eqnarray}
\label{eq:fluctQLErot} \dot{\delta X}&=&-\frac{\gamma}{2}\delta
X+\Delta \delta Y
+\sqrt{\gamma}\delta X_{a}^{\rm in},\nonumber \\
\dot{\delta Y}&=&-\frac{\gamma}{2}\delta Y-\Delta \delta X
+G\delta \phi+\sqrt{\gamma}\delta Y_{a}^{\rm in},\nonumber \\
\dot{\delta\phi}&=& \omega_{\phi}\delta L_{z},\nonumber \\
\dot{\delta L_{z}}&=& -\omega_{\phi}\delta\phi+G \delta X
-\frac{D_{\phi}}{I}\delta L_{z}+\delta\epsilon^{\rm in},
\end{eqnarray}
where $\Delta=\delta-g\phi_{s}$ is the effective cavity detuning,
$G=ga_{s}\sqrt{2}$ is the effective optorotational parameter, and
we have redefined the field fluctuations in terms of their
quadratures as
    $$
    \delta X_{a} =(\delta a+\delta a^{\dagger})/\sqrt{2},
    $$
    $$
    \delta Y_{a} =(\delta a-\delta a^{\dagger})/i\sqrt{2}.
    $$
Eq.~(\ref{eq:fluctQLErot}) can be recast compactly as
\begin{equation}
\label{eq:fluctrot} \dot{u}(t)=B u(t)+n(t),
\end{equation}
where $u(t)=(\delta \phi,\delta L_{z},\delta X_{a},\delta Y_{a})$
is the vector of fluctuations, $n(t)=(0,\delta \epsilon^{\rm
in},\sqrt{\gamma}\delta X_{a}^{\rm in},\sqrt{\gamma}\delta
Y_{a}^{\rm in})$ is the input noise vector, and
\begin{equation}
\label{eq:RHmatrix} B =
\begin{pmatrix}
0 &  \omega_{\phi}             &  0                    &       0   \\
 - \omega_{\phi}            & -D_{\phi}/I  & G   &       0   \\
0               &    0             &     -\gamma/2                     &  \Delta\\
G &    0     & -\Delta &-\gamma/2\
\end{pmatrix}.
\end{equation}

This matrix determines the dynamic stability of the physical
system, and also a measure of the entanglement between its two
subsystems, the intracavity field and the rotating mirror.

\subsection{Dynamic stability}
\label{subsec:stab}

According to the Routh-Hurwitz criterion, the stability of the
steady-state solution is assured if none of the eigenvalues of the
matrix $B$ has a positive real part. This requirement can be
reframed as a series of inequalities that have to be obeyed by the
matrix elements of $B$ \cite{dejesus1987}. These inequalities,
given by
\begin{eqnarray}
 \label{eq:RHcond}
&&D_{\phi}I^{2}\gamma\left[16\omega_{\phi}^{4}+32G^{2}\omega_{\phi}\Delta+8\omega_{\phi}^{2}(\gamma^{2}-4\Delta^{2})\right.\nonumber\\
&+&\left.(\gamma^{2}+4\Delta^{2})^{2})\right]
+16 G^{2}I^{3}\omega_{\phi}\gamma^{2}\Delta+4D_{\phi}^{3}(\gamma^{3}+4\gamma\Delta^{2}) \nonumber\\
&+&4D_{\phi}^{2}I(4\omega_{\phi}^{2}\gamma^{2}+\gamma^{4}+4G^{2}\omega_{\phi}\Delta+4\gamma^{2}\Delta^{2})>0,\nonumber\\
&&\omega_{\phi}(\gamma^{2}+4\Delta^{2})-4G^{2}\Delta >0.\nonumber \\
\end{eqnarray}
have been numerically verified to hold for the parameters of this
article.

\section{Entanglement}
\label{sec:ent}

We quantify the entanglement between the cavity mode and the
rotating end mirror in terms of the logarithmic negativity
$E_{N}$, a measure of entanglement that was proposed in
Ref.~\cite{vidal2002} to describe continuous Gaussian variables
and has been evaluated in a number of examples
\cite{serafini2004,adesso2004,vitali2007,laurat2005}. The
logarithmic negativity can be expressed in terms of the elements
of the correlation matrix
    \begin{equation}
    \label{eq:Cmatrix}
    C_{ij}=[\langle u_i(\infty)u_j(\infty)+
    u_{j}(\infty)u_{i}(\infty)\rangle]/2.
    \end{equation}
For the problem at hand, $C$ is a $4 \times 4$ matrix that can be
cast in the form
\begin{equation}
    C=
    \begin{pmatrix}
    R &  F           \\
    F^{T} &  S\\
    \end{pmatrix}
\end{equation}
where $R$, $F$ and $S$ are $2\times 2$ matrices. In terms of these
matrices the logarithmic negativity is \cite{vidal2002}
\begin{equation}
\label{eq:LNeg} E_{N}=\rm{max}[0,-ln(2\eta^{-})],
\end{equation}
where
\begin{equation}
\label{eq:etaminus}
\eta^{-}=2^{-1/2}\left(\sigma-[\sigma^{2}-4|C|]^{1/2}\right)^{1/2},
\end{equation}
and $\sigma=|R|+|S|-2|F|$, $|R|$ being the determinant of the
matrix $R$. Quantum entanglement occurs for $E_{N}>0$, i.e. for
$\eta^{-}< 1/2$.

The bipartite entanglement between the cavity mode and the
rotating mirror can  be evaluated directly from the knowledge of
the matrix $B$ \cite{vitali2007}. Integrating
Eq.~(\ref{eq:fluctrot}) formally we have
\begin{equation}
\label{eq:solvefluc}
u(t)=M(t)u(0)+\int_{0}^{t}ds M(s)n(t-s),
\end{equation}
where $M(s)=\exp(Bs)$. When the inequalities in
Eq.(\ref{eq:RHcond}) are obeyed the system is stable, as we have
seen, hence $M(\infty)=0$ so that
\begin{equation}
\label{eq:uinf}
u(\infty)=\displaystyle\lim_{t\to\infty} \int_{0}^{t}ds \,M(s)n(t-s),
\end{equation}
and the steady state is independent of the initial conditions
$u(0)$ as it should be.

The stationary quantum fluctuations are zero-mean Gaussian
processes fully characterized by the $4 \times 4$ correlation matrix
$C_{ij}$ which reads, with Eq.~(\ref{eq:uinf}),
\begin{equation}
\label{eq:corrmat}
C_{ij}=\sum_{k,l}\int_{0}^{\infty}ds \int_{0}^{\infty}ds'M_{ik}(s)M_{jl}(s')\Phi_{kl}(s-s'),
\end{equation}
where
    $$
    \Phi_{kl}(s-s')=(\langle n_{k}(s)n_{l}(s')+ n_{l}(s)n_{k}(s')\rangle)/2
    $$
characterizes the stationary noise correlations. In the limit of a
large mechanical quality factor, Eq.~(\ref{eq:Brownian2}) yields
readily $\Phi_{kl}(s-s')=D_{kl}\delta(s-s')$, where
\begin{equation}
\label{eq:Dmatrix} D =
\begin{pmatrix}
0 &  0             &  0                    &       0   \\
0          & D_{\phi}(2\bar{n}+1)/I & 0   &       0   \\
0               &    0             &     \gamma/2                     & 0\\
0 &    0     & 0 &\gamma/2\
\end{pmatrix},
\end{equation}
and Eq.~(\ref{eq:corrmat}) simplifies to
\begin{equation}
\label{eq:Cint}
C=\int_{0}^{\infty}dsM(s)DM(s)^{T},
\end{equation}
which can be integrated by parts to give
\begin{equation}
\label{eq:Lyap}
BC+CB^{T}=-D,
\end{equation}
see Appendix~\ref{sec:appa}. Using Eqs.~(\ref{eq:RHmatrix}) and
(\ref{eq:Dmatrix}), Eq.~(\ref{eq:Lyap}) can be solved analytically
to yield the correlation matrix $C$, see Appendix~\ref{sec:appa}.
The full solution is quite complicated and its form is not
particularly instructive, hence it will not be reproduced here.

We have evaluated numerically the logarithmic negativity $E_N$ for
various parameters. Our main results are summarized in
Figs.~\ref{fig:RotFieldEntpic2}-\ref{fig:RotFieldEntpic4} for the
parameter values of Table.~\ref{tab:mirrparam}, where the
mechanical quality is
\begin{equation}
\label{eq:Qphi}
Q_{\phi}=\frac{\omega_{\phi}}{D_{\phi}/I},
\end{equation}
and $F$ is the optical finesse. Our parameters are somewhat
similar to the case of the linearly vibrating mirror of
Ref.~\cite{vitali2007}, hence we make comparisons between specific
figures in that article and the present one when appropriate.

\subsection{High temperature regime}
\label{subsec:temphigh}

\begin{figure}
\includegraphics[width=0.5 \textwidth]{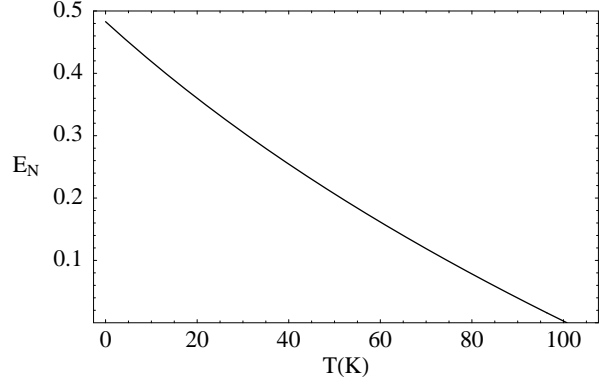}
\caption{\label{fig:RotFieldEntpic2} Logarithmic negativity
$E_{N}$ as a function of ambient temperature, for the parameters
of Table~\ref{tab:mirrparam}.}
\end{figure}

Fig.~\ref{fig:RotFieldEntpic2}
shows $E_{N}$ as a function of the temperature $T$ of the mirror
environment. Our results, which resemble those reported in
Fig.~\ref{fig:RotFieldEntpic2} of Ref.~\cite{vitali2007}, indicate
that quantum entanglement persists until roughly 100K, that is, it
should therefore be measurable in experiments in cryogenic 
environments.

We remark that the temperature in Fig.~\ref{fig:RotFieldEntpic2}
$T$ is \textit{not} the temperature of the mirror; as is well
known the presence of red-detuned laser light in the cavity can
increase the mirror damping from $D_{\phi}$ to $D_{\rm eff}$ and
thus lower the effective temperature of the mirror $T_{\rm eff}$
significantly below $T$. More precisely,
\begin{equation}
\label{eq:mirrtemp}
T_{\rm eff}=\left(\frac{D_{\phi}}{D_{\rm eff}}\right)T.
\end{equation}
For the case of $T=50K$, for example, we find $T_{\rm eff}=10\mu$K
for the parameters of Table~\ref{tab:mirrparam}, where $D_{\rm
eff}$ is (see Eq.~(7) of Ref.~\cite{mishpm2}),
\begin{eqnarray}
\label{eq:eff1} D_{\rm eff}&=&D_{\phi} + \frac{2\xi_{\phi}^{2}
\gamma P_{\rm in}}{\omega_{c}} \left (\frac{\Delta}{\Delta^{2}+
(\gamma/2)^2}\right )\nonumber
\\
&\times&\frac{\gamma}{\left[(\gamma/2)^2+(\omega-\Delta)^{2}\right]
\left[(\gamma/2)^2+(\omega+\Delta)^2\right]}.
\end{eqnarray}
Here $\omega$ is the response frequency of the system, taken in
our case to be $\omega_{\rm eff}$ [see Eq.(\ref{eq:eff})].

\subsection{Low temperature regime}
\label{subsec:templow}

\begin{figure}
\includegraphics[width=0.48 \textwidth]{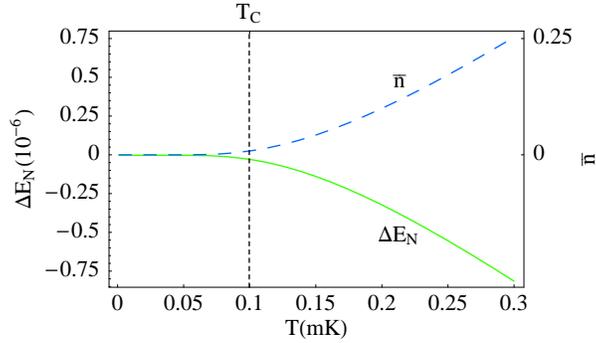}
\caption{\label{fig:RotFieldEntpic3}(Color online) Solid line:
Deviation $\Delta E_{N}$ of $E_{N}$ from its maximum value
$E_{0}\sim 0.5$ as a function of $T$. The graph shows the low
temperature end of the curve. Dashed line: mean number of thermal
phonons $\bar{n}$ [Eq.(\ref{eq:nbar})] at frequency
$\omega_{\rm eff}$ as a function of $T$. The vertical dotted line
corresponds to the temperature $T_{c}$ at which thermal phonons of
energy larger than $\hbar \omega_{\rm eff}$ start exciting the
rotor, degrading the entanglement. Parameters of
Table~\ref{tab:mirrparam}.}
\end{figure}

Figure \ref{fig:RotFieldEntpic3}
is a blown-up view of the portion of
Fig.~\ref{fig:RotFieldEntpic2} for low ambient temperatures. The
entanglement is maximum at $T \sim 0$, as would be intuitively
expected. It remains at that value as the temperature is increased
until a temperature $T_c$ where it undergoes a sharp decrease. At
that temperature the peak frequency of the thermal phonons becomes
larger than $\omega_{\rm eff}$ and they are capable of randomly
exciting the rotating mirror, thus degrading its entanglement with
the cavity mode.

This point is further illustrated in
Fig.~\ref{fig:RotFieldEntpic3}, which shows the mean number of
thermal phonons $\bar{n}$ at frequency $\omega_{\rm eff}$,
[Eq.(\ref{eq:nbar})] as a function of $T$. It increases noticeably
for temperatures above $T_{c}$ and is clearly anti-correlated to
$E_N$. This is consistent with an approximate calculation that
shows that to lowest order in $\bar{n}$ we have
\begin{equation}
 \label{eq:entanal}
E_{N} \simeq E_{0}-\kappa \bar{n},
\end{equation}
where $E_{0}$ is the maximum value of the entanglement and
$\kappa$ is a constant whose value is a function of the parameters
of the system. The analytical forms of $E_{0}$ and $\kappa$ are
involved. For the parameters of Table~\ref{tab:mirrparam} their
values are $E_{0} \sim 0.5$ and $\kappa \sim 3.1\times 10^{-6}$.

\subsection{Detuning}
\label{subsec:det}

\begin{figure}
\includegraphics[width=0.50 \textwidth]{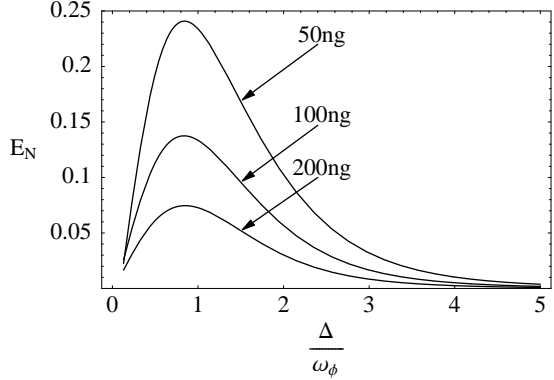}
\caption{\label{fig:RotFieldEntpic4} The logarithmic
negativity $E_{N}$ plotted as a function of the detuning
$\Delta$ (normalized by the angular frequency
$\omega_{\phi}$ of the rotating mirror) of the laser
radiation from the cavity resonance. The three different
curves correspond to different masses of the rotating
mirror. The maximum of the entanglement occurs practically
at $\Delta=\omega_{\rm eff} \sim 0.8\omega_{\phi}$ (see
the discussion in Section~\ref{subsec:det}) for all three
curves. For this plot we used the parameters from
Table~\ref{tab:mirrparam} except for the following
changes : $L=0.1$mm, $R=250 \mu$m, $Q_{\phi}=10^{7}$,
$F=10^{4}$, and $T=400$mK.}
\end{figure}

Fig.~\ref{fig:RotFieldEntpic4}
shows the logarithmic negativity as a function of the detuning
$\Delta=\omega_c-\omega_L-\phi_s =
\omega_c-\omega_L-g^2|a_s|^2/\omega_\phi$ for three masses of the
rotating mirror. This result may be compared to Fig.~1 of
Ref.\cite{vitali2007}. A careful examination of the detuning
results reveals that for each mass the entanglement is maximized
when the detuning is equal to the mirror's \textit{effective}
angular frequency of rotation $\Delta = \omega_{\rm eff}$, As is
known, the presence of red-detuned radiation $(\Delta>0)$ causes
anti-trapping and thus lowers the effective mechanical frequency
of the mirror \cite{corbitt2007}. The effective frequency
$\omega_{\rm eff}$ is given explicitly by
\begin{eqnarray}
\label{eq:eff}
    \omega_{\rm eff}^2
    &=&\omega_{\phi}^2-\frac{2\xi_{\phi}^2 \gamma P_{\rm
    in}}{I\omega_c}
    \left (\frac{\Delta}{\Delta^2+ (\gamma/2)^2}\right ) \nonumber \\
    &\times& \frac{((\gamma/2)^2-(\omega^2-\Delta^2)}
    {\left[(\gamma/2)^2+(\omega-\Delta)^{2}\right]
    \left[(\gamma/2)^2+(\omega+\Delta)^{2}\right]},
\end{eqnarray}
hence $\omega_{\rm eff}$ scales as $1/\sqrt{I}$, that is, as the
inverse square root of the mass.  Note however that for the
parameters of Fig.~\ref{fig:RotFieldEntpic4}, $\omega_{\rm eff}
\sim 0.8 \omega_{\phi}$ for all choices of mass. Thus the maximum
entanglement occurs for $\Delta/\omega_{\phi}\sim 0.8.$)

Physically, the condition $\Delta = \omega_{\rm eff}$ corresponds
to the optimum detuning for optomechanical cooling
\cite{marquardt2007} and can be understood in terms of cavity
enhanced scattering of the anti-Stokes photons by the rotating
mirror \cite{schliesser2006}. Thus it is not surprising that
$E_{N}$ reaches a maximum at this detuning.

\subsection{Angular momentum}
\label{subsec:am}

\begin{figure}
\includegraphics[width=0.50 \textwidth]{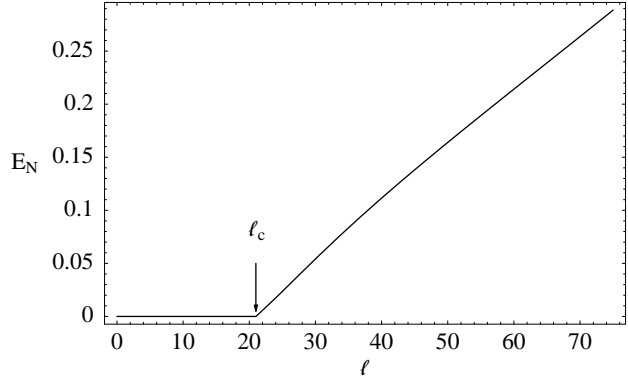}
\caption{\label{fig:RotFieldEntpic5} Logarithmic negativity
$E_{N}$ as a function of the orbital angular momentum carried by
the Laguerre-Gaussian mode. The threshold angular momentum $l_{c}
\sim 21$ at which entanglement begins to appear is indicated.
Parameters of Table~\ref{tab:mirrparam}, temperature $T=10$K.}
\end{figure}

As pointed out in Ref.~\cite{mishpm2}, in the case of optical
cooling of a vibrating mirror the linear momentum of the photons
is difficult to change experimentally, and essentially entails
replacement of the laser radiation source. For example to increase
the linear momentum just by a factor of two would require an
ultraviolet laser in place of an infrared laser. In contrast in
the case of rotational cooling it is comparatively easier to
achieve higher $l$. In light of this fact we also studied the
entanglement as a function of the orbital angular momentum carried
away from the mirror by the Laguerre-Gaussian mode,. The results
of this analysis are summarized in Fig.~\ref{fig:RotFieldEntpic5}.
For the temperature $T=10$K considered in the figure, entanglement
appears only beyond the threshold value $l_{c}\sim 21$. This is
consistent with Eq.~(\ref{eq:optrotcon}) which shows that for
sufficiently low angular momentum, i.e. $l<l_{c}$, the
optorotational coupling between radiation and the rotating mirror
may not be high enough for entanglement to occur. The value of
$l_{c}$ depends on the other parameters that contribute to the
optorotational coupling and of course also on the ambient
temperature $T$. In general we found that entanglement appears
when the effects of radiation torque coupling $g$ are strong
enough to counter the thermal noise, which increases perceptibly
with temperature beyond $T=T_c.$

\begin{table*}
\begin{center}
\renewcommand{\arraystretch}{1.4}
\begin{tabular*}{0.7\textwidth}{|c|c|c|c|c|}
\hline \textbf{No.} & \textbf{Parameter} & \textbf{Description} & \textbf{Value} & \textbf{Units} \\
\hline
\hline
1.                  & $L$                & Cavity length                                   & 1                  &  mm         \\
\hline
2.                  & $\lambda$          & Laser wavelength                                & 810                &  nm         \\
\hline
3.                  & $\omega_{c}$       & Cavity resonance frequency                      & $2\pi$10$^{14}$    &  Hz         \\
\hline
4.                 & $\omega_{\phi}$     & Rotating mirror angular frequency               & 2$\pi$10           &  MHz        \\
\hline
5.                  & $M$                & Rotating mirror mass                            & 100                &  ng         \\
\hline
5.                 & $R$                 & Rotating mirror radius                          & 10                 &  $\mu$m      \\
\hline
7.                 & $Q_{\phi}$          & Mechanical quality factor                       & 2$\times 10^{6}$   &     -         \\
\hline
8.                 & $F$                 & Optical finesse                                 & 5$\times 10^{3}$   &   -           \\
\hline
9.                 & $l$                 & Orbital angular momentum                        &   100              &   $\hbar$      \\
\hline
10.                 & $P_{\rm in}$       & Input laser power                               &   50               &   mW         \\
\hline
11.                 & $\Delta/\omega_{\phi}$           & Normalized laser detuning         &   1                &    -         \\
\hline
\hline
\end{tabular*}
\end{center}
\caption{\label{tab:mirrparam} Definitions and approximate
values of some of the parameters used in the text.}
\end{table*}

\section{Conclusion}
\label{sec:conc}

To conclude, we have demonstrated theoretically that radiation
torque can serve to entangle a Laguerre-Gaussian cavity field with
a rotating spiral phase element. We have shown that the physical
basis of entanglement is the exchange of angular momentum between
the cavity mode and the rotating mirror. The mirror-field
entanglement was investigated as a function of temperature and
shown to be measurable at the temperatures routinely achieved by
cryogenic apparatus.

The entanglement was also shown to remain at its maximum value
until a critical temperature at which the peak of the thermal
spectrum occurred at the frequency corresponding to the rotor
quanta, and beyond which the random excitation due to the thermal
photons degraded the entanglement. The behavior of the
entanglement with respect to cavity detuning was shown to be
consistent with the mechanism of optomechanical cooling, i.e. the
entanglement was maximum at a detuning known to correspond to
optimal cooling of the mirror. Further, the fact that the angular
momentum of the light can be changed relatively easily justified
examining the entanglement with respect to changes in angular
momentum. The main result here is that the mirror-radiation
coupling only becomes strong enough to overcome the thermal noise
and yield entanglement beyond a critical value of the angular
momentum.

The measurement of the entanglement investigated in this paper can
be carried out in a manner analogous to that proposed in the case
of a linearly vibrating cavity and illustrated in Fig. 3 of
Ref.~\cite{vitali2007}. In the present case the mirrors are
replaced by spiral phase elements, and a second optical cavity can
be formed by placing a third transmissive spiral phase element
beyond the rotating mirror of Fig.~\ref{fig:RotFieldEntpic1}. By
using a weak intracavity field for this second cavity all the
elements of the correlation matrix $C$ can be obtained as proposed
in Ref.~\cite{vitali2007} and demonstrated in a measurement of the
entanglement between two optical fields \cite{laurat2005}.

\begin{acknowledgments}
This work is supported in part by the US Office of Naval
Research, by the National Science Foundation and by the
US Army Research Office. We acknowledge fruitful
discussions with H. Uys, O. Dutta and M. Leclerc.
\end{acknowledgments}
\appendix
\section{Lyapunov Equation}
\label{sec:appa}

Here we derive Eq.(\ref{eq:Lyap}), which is a typical
Lyapunov equation encountered in the study of dynamical systems.
Also, we show our method for solving the equation.

\subsection{Derivation}
\label{subsec:deriv}

Beginning from Eq.(\ref{eq:Cint})
\begin{equation}
\label{eq:C}
C=\int_{0}^{\infty}dsM(s)DM(s)^{T},
\end{equation}
where $M(s)=e^{Bs}$ and the matrices $B,D$ are given
by Eqs.(\ref{eq:RHmatrix}) and (\ref{eq:Dmatrix})
respectively, we integrate by parts using
$f'(s)=e^{Bs}$ and $g(s)=De^{B^{T}s}$ in the rule
\begin{equation}
\label{eq:intparts}
\int_{0}^{\infty} ds \,f'(s)g(s)= \left[f(s)g(s)\right]_{0}^{\infty} -\int_{0}^{\infty} ds \,f(s)g'(s).
\end{equation}
This yields the equation
\begin{eqnarray}
\label{eq:Cintparts}
\begin{array}{lcl}
C&=&  \left[B^{-1} e^{Bs} De^{B^{T}s}\right]_{0}^{\infty}-\displaystyle\int_{0}^{\infty}ds \, B^{-1}e^{Bs}De^{B^{T}s}B^{T} \\
 &=& -B^{-1}D-B^{-1}(\displaystyle\int_{0}^{\infty} ds \,e^{Bs}De^{B^{T}s})B^{T}\\
 &=& -B^{-1}D-B^{-1}CB^{T},\\
\end{array}
\end{eqnarray}
where in the second step we have used the fact if the
Routh-Hurwitz criterion is satisfied then
$\displaystyle\lim_{s\to\infty} e^{Bs}=0$, since
all the eigenvalues of $B$ have negative real parts.
We then pre-multiply the last line of
Eq.~(\ref{eq:Cintparts}) by $B$ and rearrange to obtain
\begin{equation}
\label{eq:Lyapapp}
BC+CB^{T}=-D,
\end{equation}
which is Eq.~(\ref{eq:Lyap}).

\subsection{Solution}
\label{subsec:solve}

Here we record our method of solving Eq.~(\ref{eq:Lyapapp})
for the correlation matrix $C$. By definition [see above
Eq.~(\ref{eq:corrmat})] $C$ is a $4\times 4$ real symmetric
matrix. Thus we can write
\begin{equation}
\label{eq:C1}
C=\left(
\begin{array}{llll}
 \lambda _1 & a_1 & a_2 & a_3 \\
 a_1 & \lambda _2 & b_1 & b_2 \\
 a_2 & b_1 & \lambda _3 & c_1 \\
 a_3 & b_2 & c_1 & \lambda _4
\end{array}
\right)
\end{equation}
where we have denoted the diagonal elements with Greek
letters and the off-diagonal with Roman letters. We now
define a matrix $C'$ where
\begin{equation}
\label{eq:cprime}
C'=BC+CB^{T}+D,
\end{equation}
which equals zero as per Eq.~(\ref{eq:Lyap}). Evaluating
the RHS of Eq.~(\ref{eq:cprime}) we find that at least
one matrix element of $C'$ involves a single non-diagonal
element of $C$. This observation, which is central to our
method, turns out to be true at every iteration, where the
process of iteration is defined below.

We now solve for the non-diagonal element of $C$ mentioned
above by equating the relevant matrix element of $C'$ to zero.
We then use that value of the non-diagonal element
in $C$ and re-evaluate $C'$. This process is performed
iteratively until all the off-diagonal elements of $C$
are expressed in terms of the diagonal elements of
$C$. At this stage $C$ has the form
\begin{widetext}
\begin{equation}
 \label{eq:cfinal} C=
\left(
\begin{array}{llll}
 \lambda _1 & 0 & \frac{\left(\lambda _1-\lambda _2\right) \omega _{\phi }}{G} & \frac{\gamma  \left(\lambda _3+\lambda _4-1\right)}{2 G} \\
 0 & \lambda _2 & -\frac{\gamma _{\phi } \left(2 \bar{n}-2 \lambda _2+1\right)}{2 G} &
\frac{\gamma ^2(\lambda_{3}+\lambda _4-1)-4 \Delta  \lambda
   _2 \omega _{\phi }-4 \lambda _1 \left(G^2-\Delta  \omega _{\phi }\right)}{4 G \omega _{\phi }} \\
 \frac{\left(\lambda _1-\lambda _2\right) \omega _{\phi }}{G} & -\frac{\gamma _{\phi }
\left(2 \bar{n}-2 \lambda _2+1\right)}{2 G} & \lambda _3 & \frac{\gamma
   \left(2 \lambda _3-1\right)}{4 \Delta } \\
 \frac{\gamma  \left(\lambda _3+\lambda _4-1\right)}{2 G} &
\frac{\gamma^2(\lambda _3 +\lambda _4 -1)-4 \Delta  \lambda _2 \omega _{\phi }-4 \lambda
   _1 \left(G^2-\Delta  \omega _{\phi }\right)}{4 G \omega _{\phi }} & \frac{\gamma  \left(2 \lambda _3-1\right)}{4 \Delta } & \lambda _4
\end{array}
\right).
\end{equation}
\end{widetext}
Also we find $C'=0$ yields 4 equations in the diagonal
elements $\lambda_{1,2,3,4}$ of $C$, and these can be easily
solved, giving us analytical expressions for all the elements
of the $C$ matrix.

\end{document}